\documentclass{PoS}

\newcommand{\tn}[1]{\textnormal{#1}}
\newcommand{\wQ}{\it Q}
\newcommand{\ee}{e^+ e^-}
\newcommand{\pp}{\pi^+ \pi^-}
\newcommand{\kk}{K^+ K^-}
\newcommand{\ep}{\eta \pi^0}
\newcommand{\upp}{\Upsilon(1S) \pi^+ \pi^-}
\newcommand{\ukk}{\Upsilon(1S) K^+ K^-}
\newcommand{\uep}{\Upsilon(1S) \eta \pi^0}
\newcommand{\uos}{\Upsilon(1S)}

\newcommand{\yb}{Y_{b}}

\title{Theory Overview on Spectroscopy}

\ShortTitle{Spectroscopy-Overview}

\author{\speaker{Ahmed Ali}\\
        Deutsches-Elektronen Synchrotron DESY, Notkestrasse 85, D-22607 Hamburg\\
        E-mail: \email{ahmed.ali@desy.de}}

\abstract{A theoretical overview of the exotic spectroscopy in the charm and beauty quark
sector is presented. These states are unexpected harvest from the $e^+e^-$ and hadron colliders
and a permanent abode for the majority of them has yet to be found. We argue that some of these
states, in particular the $Y_b(10890)$ and the recently discovered states $Z_b(10610)$
and $Z_b(10650)$, discovered by the Belle collaboration are excellent candidates for tetraquark states
 $[bq][\bar{b}\bar{q}]$, with $q=u,d$ light quarks. Theoretical analyes of the Belle data carried out
in the tetraquark context is reviewed.}

\FullConference{The 13th International Conference on B-Physics at Hadron Machines - Beauty2011,\\
		April 04-08, 2011\\
		Amsterdam, The Netherlands}

\begin{document}

\section{Introduction}

The title of my talk is both ambitious and pretentious!
I hasten to state that the 
mandate given to me is rather limited, namely to review the phenomenology of
hadronic states discovered recently  in the mass region of the charmonia
and the bottomonia. Spearheaded by the experiments at the B factories
and the Tevatron, with the experiments at the LHC as welcome new-comers,
an impressive number of new states have been reported. Generically called $X$, $Y$ and $Z$,
these states defy
a conventional quarkonia interpretation; this certainly holds for  the majority of them.
Their gross properties, such as the spin-parity assignments, masses, production mechanisms
and decay modes, have been discussed in a number of comprehensive
reviews~\cite{Olsen:2009gi,Brambilla:2010cs}.

 There have been
a number of more recent developments in the field of quarkonium spectroscopy and I will
confine myself just to their discussion. They involve the observation of the two charged
bottomonium-like resonances by the Belle Collaboration~\cite{Collaboration:2011gj}
in the $\pi^\pm \Upsilon(nS)~(n=1,2,3)$ and $\pi^\pm h_b(mP)~(m=1,2)$ mass spectra that are
produced in association with a single charged pion in $e^+e^-$ annihilation at energies
 near the $\Upsilon(5S)$ resonance. Here $h_b(mP)$ are the P-wave spin-singlet  bottomonia
states.
Calling the charged particles $Z_b(10610)$ and $Z_b(10650)$, their masses and the decay widths
 averaged over the five final states are, respectively,  $M[Z_b(10610)]=10608.4 \pm 2.0$ MeV,
$\Gamma[Z_b(10610)]= 15.6 \pm 2.5$ MeV, and  $M[Z_b(10650)]=10653.2 \pm 1.5$ MeV,
$\Gamma[Z_b(10650)]= 14.4 \pm 3.2$ MeV. The favoured quantum number assignments for both are
$I^G(J^P)=1^+(1^+)$. This discovery was preceded by the observation of the  $h_b(1P)$ 
and $h_b(2P)$ states, also by the Belle Collaboration~\cite{Adachi:2011ji} in the
reaction $e^+e^- \to h_b(nP)\pi^+\pi^-$, with the masses
$M[h_b(1P)]= (9898.25 \pm 1.06 ^{+1.03}_{-1.07})$ MeV and
$M[h_b(1P)]= (10259.76 \pm 0.64 ^{+1.43}_{-1.03})$ MeV. These measurements yield
hyperfine splitting in the bottomonium sector, defined as the mass difference between the
 $P$-wave spin-singlet
state $h_b(mP)$ and the 
weighted average of the corresponding $P$-wave triplet states, $\chi_{bJ}(nP)$,
$\Delta M_{\rm HF}(nP)\equiv \langle M(n^3P_J)\rangle - M(n^1P_1)$, with
 $\Delta M_{\rm HF}(1P)=(1.62\pm 1.52)$ MeV and $\Delta M_{\rm HF}(2P)=(0.48^{+1.57}_{-1.22})$ MeV.
They are consistent with theoretical expectations and also with the hyperfine splitting measured
in the charmonium sector $\Delta M_{\rm HF}=(0.14 \pm 0.30)$ MeV~\cite{Amsler:2008zzb},
consistent with zero. Theoretically expected widths
 of $h_b(1P)$ and $h_b(2P)$ are of order 100 keV~\cite{Godfrey:2002rp}, which are too
small to be measured by Belle.

Still on the subject of $h_b(1P)$, the BaBar collaboration~\cite{:2011zp} has presented
 evidence of its production in the decay $\Upsilon(3S) \to \pi^0 h_b(1P)$, followed by the
decay $h_b(1P) \to \gamma \eta_b(1S)$, in the distribution of the recoil mass
against the $\pi^0$ at the mass $M[h_b(1P)]=(9902 \pm 4 \pm 1)$ MeV, which is
consistent with the Belle measurements~\cite{Adachi:2011ji}. The width of $h_b(1P)$
is consistent with the experimental resolution, and the reported product branching ratio is
${\cal B}(\Upsilon(3S) \to \pi^0h_b) \times {\cal B}(h_b \to \gamma \eta_b)= (3.7 \pm 1.1 \pm 0.7)\times 10^{-4}$. In this, and also in $M[h_b(1P)]$, the first error is statistical and the
second systematic. The isospin-violating decay $\Upsilon(3S) \to \pi^0 h_b(1P)$ is
expected to have a branching fraction of about $10^{-3}$~\cite{Voloshin:1985em,Godfrey:2005un},
and the branching fraction
 ${\cal B}(h_b(1P) \to \gamma \eta_b(1S)) \sim (40 {\rm - } 50)\%$~\cite{Godfrey:2002rp};
hence, the measured product branching ratio is as anticipated theoretically.
 It is noteworthy that the
decay $\Upsilon(3S) \to h_b(1P) \pi^+\pi^-$, which is 
suppressed by at least an order of magnitude compared to the decay 
 $\Upsilon(3S) \to \pi^0 h_b(1P)$~\cite{Voloshin:1985em}, has not been observed.
The observation of the singlet $P$-state in the charmonium sector $h_c(1P)$
has also been reported this year by the CLEO collaboration~\cite{:2011uqa}
in the process $e^+e^- \to \pi^+\pi^- h_c(1P)$
at the center-of-mass energy $E_{c.m.}=4170$ MeV. In fact, CLEO pioneered the technique
of searching for peaks in the mass spectrum recoiling against the $\pi^0$, and the
resulting mass
 $M[h_c(1P)]=(3525.27 \pm 0.17)$ MeV measured by this method 
is consistent with an earlier measurement of the $h_c(1P)$ mass
from the decay $\psi(2S) \to \pi^0h_c$~\cite{Dobbs:2008ec}. The product branching ratio
${\cal B}(\psi(2S) \to \pi^0h_c) \times {\cal B}(h_c \to \gamma \eta_c)=(4.19 \pm 0.32 \pm 0.45) \times 10^{-4}$ is in agreement with theoretical expectations,
and is also very similar to what has been reported by Babar for the
corresponding  $h_b(1P)$ product branching ratio, quoted above. However, there is an intriguing
hint in the CLEO measurements of the cross section for $e^+e^- \to h_c(1P) \pi^+\pi^-$, which
rises at $E_{c.m.}=4260$ MeV. Since this is close to the mass of the $J^{PC}=1^{--}$ hadron
$Y(4260)$, which is a candidate for the hidden $c\bar{c}$ tetraquark state, it would suggest 
that the mechanism $e^+ e^- \to Y(4260) \to h_c(1P) \pi^+\pi^-$
has something to do with the rise in the
cross section. This remains to be confirmed in the next round of precise experiments.

\section{Current experimental anomalies}%

\begin{figure}[t]
\centering
\resizebox{0.95\textwidth}{0.25\textwidth}{
\includegraphics[width=0.8\textwidth,height=10cm]{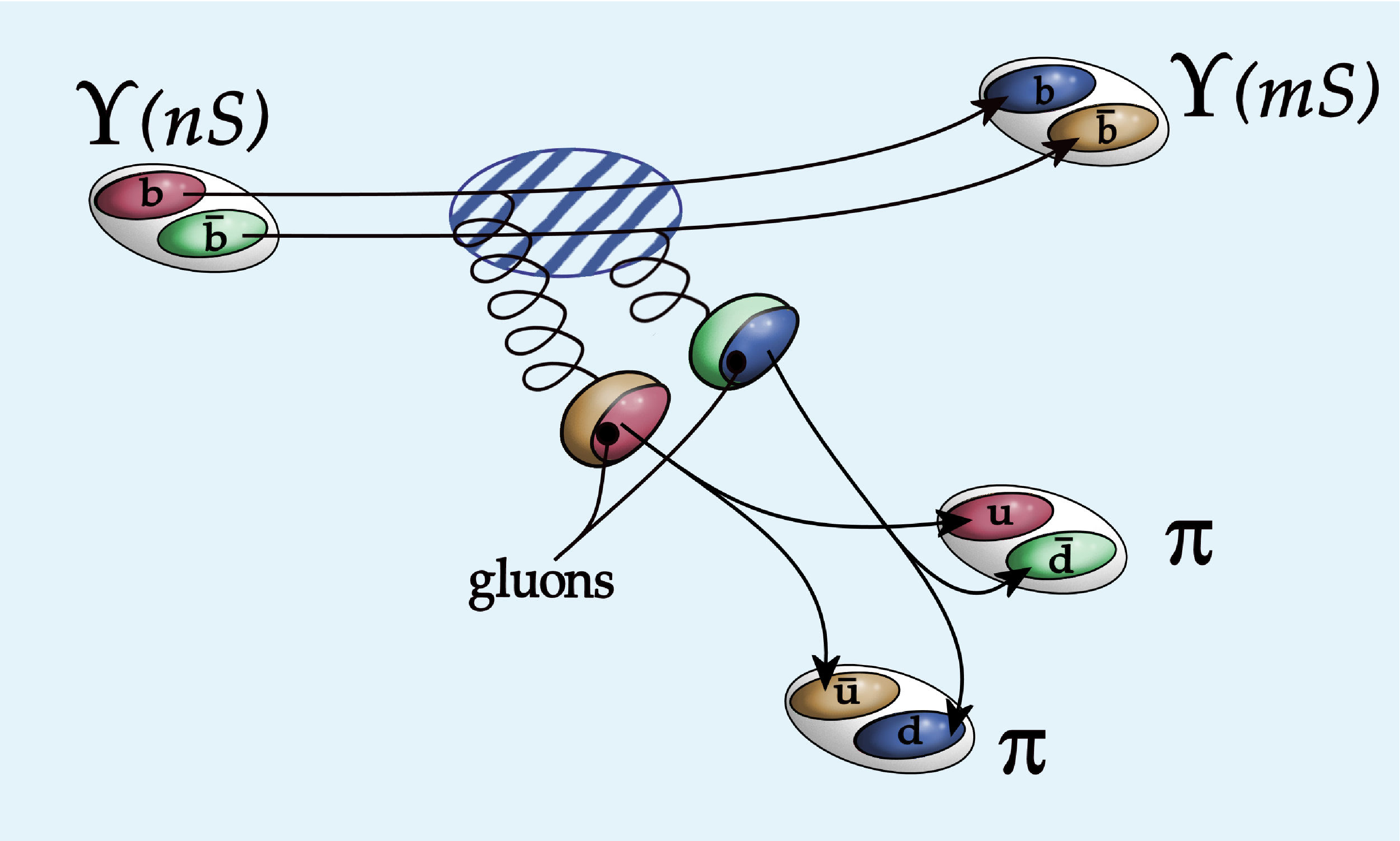}
\includegraphics[width=0.8\textwidth,height=10cm]{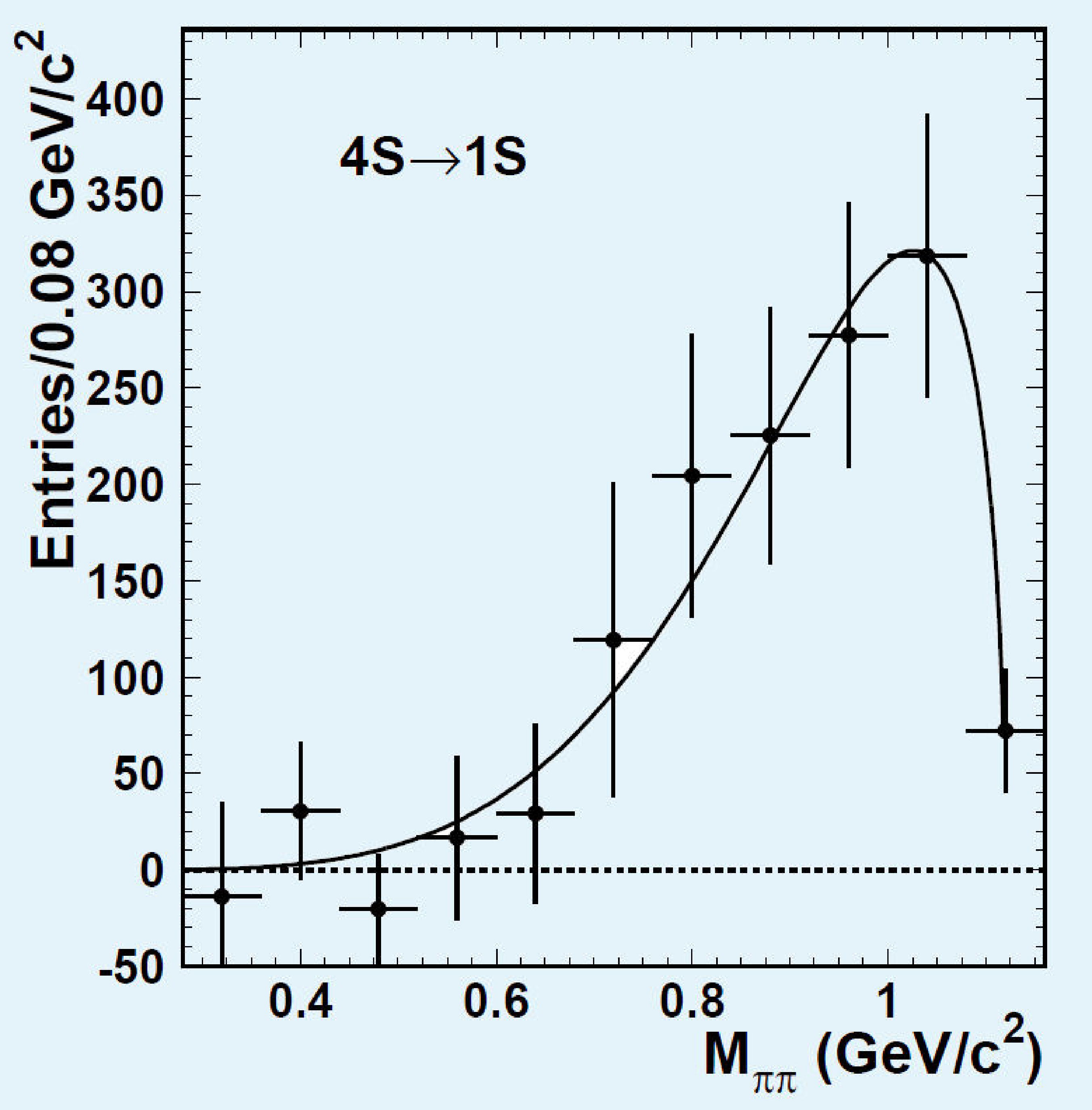}
}
\vspace*{-3mm}
\caption{Left frame: Zweig-suppressed diagram for the transition $\Upsilon(nS) \to \Upsilon(mS) \pi \pi$ with
$m < n$, which forms the basis of the QCD estimates of the decay rates and distributions in
heavy quarkonia dipionic transitions.
Right frame: The dipion invariant mass spectrum $M_{\pi \pi}$ measured in the decay
 $\Upsilon(4S) \to \Upsilon(1S) \pi \pi$ by the Belle collaboration  together with
a theoretical curve  based essentially on the diagram shown in the left frame.
 (From ~\cite{:2009zy}.) 
\label{fig:upsilon4s}
}
\end{figure}

\begin{figure}[t]
\centering
\resizebox{0.95\textwidth}{0.25\textwidth}{
\includegraphics[width=0.8\textwidth,height=10cm]{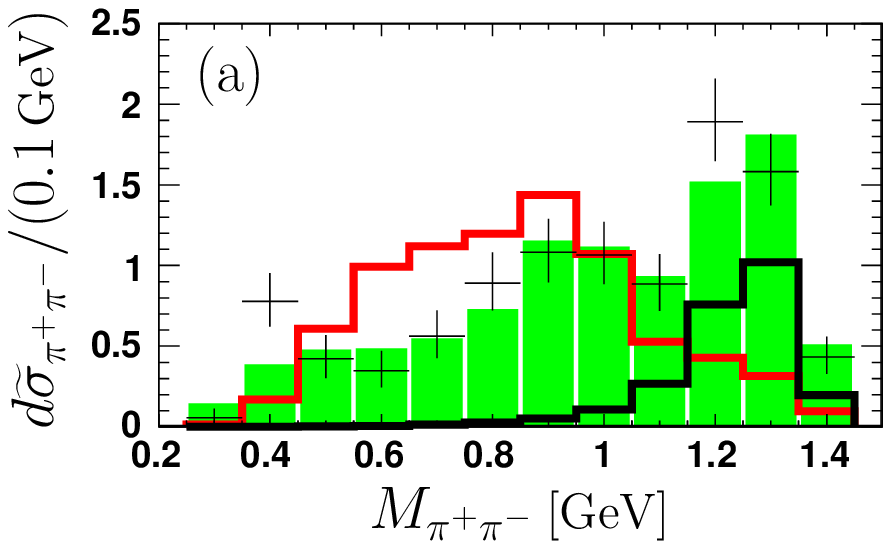}
\includegraphics[width=0.8\textwidth,height=10cm]{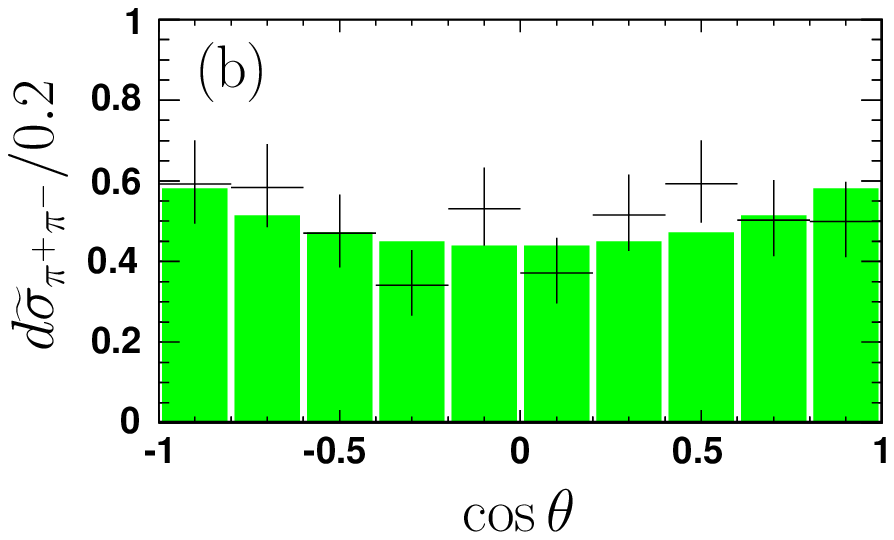}
}
\vspace*{-3mm}
\caption{Fit results of the $M_{\pi^+\pi^-}$ distribution (a)  and  the
  $\cos\theta$ distribution (b) for $e^+ e^- \to Y_b \to \Upsilon(1S) \pi^+\pi^-$, 
 normalized by the measured cross section by  Belle~\cite{Abe:2007tk}. 
 The histograms 
  represent theoretical fit results based on the tetraquarks hypothesis, while the crosses are the 
  Belle data. 
The solid curves in (a)  show purely resonant contributions from the $f_0(980)$ and $f_2(1270)$.
 (From ~\cite{Ali:2010pq}.) 
\label{fig:spectra}
}
\end{figure}

There is a number of anomalous features in the Belle 
 data taken in the center-of-mass energy region near the $\Upsilon(5S)$
mass. The first of these was reported some three years ago~\cite{Abe:2007tk,:2008pu} in the processes
$e^+ e^- \to \Upsilon(1S) \pi^+\pi^-, ~\Upsilon(2S) \pi^+\pi^-~, \Upsilon(3S) \pi^+\pi^-$, measured
in the center-of-mass energy range between 10.83 GeV and 11.02 GeV. The enigmatic features
of the Belle data are (i) the anomalously large decay widths (or cross sections) for the mentioned final states,
and (ii) the dipion invariant mass distributions recoiling against the $\Upsilon(1S)$ and $\Upsilon(2S)$
states, which are at variance with similar spectra measured in the transitions involving lower mass
bottomonium states $\Upsilon(nS) \to \Upsilon(mS) \pi^+\pi^-$ (with $m < n$).
To quantify the problem, the reported partial widths are
 $\Gamma[\Upsilon(1S)\pi^+\pi^-)]= 0.59 \pm 0.04 \pm 0.09$ MeV and 
 $\Gamma[\Upsilon(2S)\pi^+\pi^-)]= 0.85 \pm 0.07 \pm 0.16$ MeV. Compared to the
corresponding  
partial decay widths of the lower three $\Upsilon(nS)$ $(n=2,3,4)$ states,
 $\Gamma[\Upsilon(2S) \to \Upsilon(1S)\pi^+\pi^-)] \sim 6$ keV,
$\Gamma[\Upsilon(3S) \to \Upsilon(2S)\pi^+\pi^-)] \sim 0.9$ keV, and
$\Gamma[\Upsilon(4S) \to \Upsilon(1S)\pi^+\pi^-)] \sim 1.9$ keV, the production of the
$\Upsilon(nS) \pi^+\pi^-$ in the energy region near the $\Upsilon(5S)$ is larger by two to three
orders of magnitude. The order keV partial widths are well-accounted for in the QCD multipole
 expansion~\cite{Brown:1975dz,Gottfried:1977gp} 
based essentially on the Zweig-suppressed process shown in Fig.~\ref{fig:upsilon4s} (left-hand frame).
 The dipion invariant mass spectrum anticipated in the QCD multipole
 expansion is shown on the example of the decay $\Upsilon(4S) \to \Upsilon(1S) \pi^+\pi^-$ in
 Fig.~\ref{fig:upsilon4s} (right-hand frame) and compared with the data taken from the
 Belle collaboration at $\Upsilon(4S)$~\cite{:2009zy}. They are in excellent agreement with each other.
 Not so, for
the dipionic transitions measured in the $\Upsilon(5S)$ region, in which the dipionic mass
spectra are dominated by the scalar
meson  $f_0(980)$ and the tensor meson $f_2(1270)$ (for the  
$\Upsilon(1S)\pi^+\pi^-$ mode) and by the $f_0(600)$ and $f_0(980)$ mesons 
(for the $\Upsilon(2S)\pi^+\pi^-$ mode). This is illustrated in Fig.~\ref{fig:spectra} for the
process $e^+e^- \to \Upsilon(1S) \pi^+\pi^-$ which shows the distributions in the $M_{\pi^+\pi^-}$
(left-hand frame) and in the helicity angle ($\cos \theta$ distribution  (right-hand frame).
The dipion mass spectrum measured near the $\Upsilon(5S)$ clearly shows peaks at $f_0(980)$ and
 $f_2(1270)$. An 
interpretation of the process in terms of the production and decay of a $J^{PC}=1^{--}$ tetraquark state
\cite{Ali:2009es,Ali:2010pq} (histograms and the solid curves)
accounts well the experimental distributions.
 We will return to discuss the underlying dynamical model later in section 4 of this report.

Not only are the cross sections for $e^+ e^- \to \Upsilon(nS) \pi^+\pi^-$ ($n=1,2,3$)
near the $\Upsilon(5S)$ anomalously large by at least two orders of magnitude, the same holds for the
production of the P-wave spin-singlet bottomonia states $h_b(mP)$ ($m=1,2$), for which the production
cross sections for $e^+ e^- \to h_b(1P) \pi^+\pi^-$ and $e^+ e^- \to h_b(2P) \pi^+\pi^-$ are
also anomalously large~\cite{Adachi:2011ji}. 
The ratios of the production cross-sections in the indicated final states relative to that for the $e^+ e^- \to \Upsilon(2S) \pi^+\pi^-$
production are as follows~\cite{Adachi:2011ji}:
\begin{eqnarray}
\tilde{\sigma}[\Upsilon(1S) \pi^+\pi^-] &=& 0.638 \pm 0.065 ^{+0.037}_{-0.056}\nonumber\\
\tilde{\sigma}[\Upsilon(3S) \pi^+\pi^-] &=& 0.517 \pm 0.082  \pm 0.070 \nonumber\\
\tilde{\sigma}[h_b(1P) \pi^+\pi^-] &=& 0.407 \pm 0.07 ^{+0.043}_{-0.076}  \nonumber\\
\tilde{\sigma}[h_b(2P) \pi^+\pi^-] &=& 0.78 \pm 0.09 ^{+0.22}_{-0.10} 
\end{eqnarray}
We have already commented on the anomalous production cross sections in the $\Upsilon(ns) \pi^+\pi^-$
modes near the $\Upsilon(5S)$ region. The ratios given in the last two equations above for the
$h_b(1P) \pi^+\pi^-$ and $h_b(2P)\pi^+\pi^-$ are found to be of order unity, a feature which violates theoretical
expectations as the processes $\Upsilon(5S) \to h_b(mP) \pi^+\pi^-$ involve heavy quark spin-flip,
 which are suppressed  by $1/m_b$ in the amplitude. It is obvious that the production mechanisms of all five processes
involving $\Upsilon(nS)\pi^+\pi^-$ ($n=1,2,3$) and $h_b(mP) \pi^+\pi^-$ ($ m=1,2$) are exotic. In
particular, the true mechanisms at work avoid the Zweig-suppression seen in similar dipionic transitions and
 evade power suppression due to the spin-flip transitions for the $h_b(mP)\pi^+\pi^-$ case.
It is worth recalling that no excess of the kind seen in the Belle
 measurements near the $\Upsilon(5S)$~\cite{Abe:2007tk,:2008pu,Adachi:2011ji} is seen by them or
any other experiment either at energies below or above the $\Upsilon(5S)$ region. Any plausible
theoretical explanation must  account for all these features. 

These measurements have invoked a number of theoretical ideas. Particularly interesting is the
suggestion by Bondar {\it et al.}~\cite{Bondar:2011ev}, in which the resonances $Z_b(10610)$ and
$Z_b(10650)$ are assumed mostly of a 'molecular' type due to their
respective  proximity with the $B^* \bar{B}$ and
$B^*\bar{B}^*$ thresholds. Thus, the internal dynamics of the states  $Z_b(10610)$ and $Z_b(10650)$ is 
dominated by the coupling to meson pairs $B^* \bar{B}- B \bar{B}^*$ and $B^*\bar{B}^*$, respectively.
In particular, the $b\bar{b}$ pair within the $Z_b(10610)$ and $Z_b(10650)$ is an equal mixture of a
spin-triplet and spin-singlet with the relative phase orthogonal between the two resonances, i.e.,
\begin{eqnarray}
|Z_b(10610)\rangle &=&\frac{1}{\sqrt{2}} \left( 0^-_{b\bar{b}} \otimes 1^-_{\bar{Q}q} -
                                                1^-_{b\bar{b}} \otimes 0^-_{\bar{Q}q}\right)~,\nonumber\\
|Z_b(10650)\rangle &=&\frac{1}{\sqrt{2}} \left( 0^-_{b\bar{b}} \otimes 1^-_{\bar{Q}q} +
                                                1^-_{b\bar{b}} \otimes 0^-_{\bar{Q}q}\right)~.
\label{eq:bonder-eq}
\end{eqnarray}
Here $0^-$ and $1^-$ stand for the para- and ortho-states with negative parity. 
The assignments (\ref{eq:bonder-eq}) would predict that the mass difference $M[Z_b(10650)] -M[Z_b(10610)]$
should be equal to that between the $B$ and $B^*$ masses. The observed mass difference of 46
MeV~\cite{Adachi:2011ji} is in neat agreement with this argument. The spin-structure in
 (\ref{eq:bonder-eq}) also suggests that the resonances $Z_b(10610)$ and $Z_b(10650)$ have the same decay
 width. This again is in agreement within measurement errors with the Belle data~\cite{Adachi:2011ji}:
 $\Gamma[Z_b(10610)]=15.6 \pm 2.5$ MeV and  $\Gamma[Z_b(10650)]= 14.4 \pm  3.2$ MeV. The maximal 
ortho-para mixing of the heavy quarks in the $Z_b(10610)$ and $Z_b(10650)$ resonances described by
Eq.~(\ref{eq:bonder-eq}) also  implies  couplings of comparable strengths to channels with states of ortho-
and para-bottomonium, leading to the following couplings of these resonances to the channels $\Upsilon(nS) \pi^\pm$
and $h_b(mP)\pi^\pm$\cite{Bondar:2011ev}:
\begin{equation}
C_h \, \, E_{\pi}\, \vec{\Upsilon}(nS) \cdot (\vec{Z}_b(10610) -\vec{Z}_b(10650))~, ~~~~
C_{\Upsilon} (\vec{p}_\pi\,\, \times \vec{h}_b)\cdot (\vec{Z}_b(10610) + \vec{Z}_b(10650))~,
\label{eq:bonder-eq4}
\end{equation} 
where $\vec{Z}_b(10610)$, $\vec{Z}_b(10650)$ and $\vec{h}_b$ denote the polarization vectors of the
corresponding spin-1 states, and $E_\pi$ and $\vec{p}_\pi$ are the pion energy and its three-momentum,
respectively; $C_h$ and $C_{\Upsilon}$ are {\it a priori} unknown coupling constants to be determined by
data. The amplitudes described by Eq.~(\ref{eq:bonder-eq4})  applied to the decays
 $\Upsilon(5S) \to \Upsilon(nS)\pi^+\pi^-$ and $\Upsilon(5S) \to h_b(mS)\pi^+\pi^-$ yield the right
pattern of destructive and constructive interferences seen in the Dalitz distributions of these
processes~\cite{Adachi:2011ji}. All of these arguments are plausible. Further variations on the molecular
 theme and predictions can be seen in
 ~\cite{Voloshin:2011qa,Cleven:2011gp,Yang:2011rp,Sun:2011uh}.

 However, the structure suggested in Eq.~(\ref{eq:bonder-eq}) is a postulate not yet seen  in
decays other than those of the $\Upsilon(5S)$. A particular case in point are the decays of
the $\Upsilon(6S)$, where the available phase space for the decays
 $\Upsilon(6S) \to \Upsilon(nS) \pi^+\pi^-$ and $\Upsilon(6S) \to h_b(mP) \pi^+\pi^-$ are much
larger. Hence, the implications of Eqs.~(\ref{eq:bonder-eq}) and (\ref{eq:bonder-eq4}) should be, at
least qualitatively, very similar to those discussed in the context of the Belle data from the
$\Upsilon(5S)$ region. This remains to be tested. In addition, there are also
some specific features of the Belle data which do not go hand-in-hand with the usual understanding of a hadronic molecule,
the closest example of which is the Deuteron. The masses  of the $Z_b(10610)$ and $Z_b(10650)$
are above the respective thresholds. The Deuteron mass, on the other hand,
lies below  the threshold by about 2.2 MeV. Also, the decay widths of the $Z_b(10610)$ and $Z_b(10650)$
are not particularly small, as one would expect for a hadron molecule.
 On the contrary, their decay widths are similar in order of magnitude
as that of the $\Upsilon(5S)$.
This is also curious as the other 'hadronic molecule' discussed at length in a similar context,
namely the $X(3872)$, has a much smaller (by at least an order of magnitude) decay width, with
the current 90\% C.L.~limit being $\Gamma[X(3872)] < 1.2$ MeV~\cite{Choi:2011fc}. 

In the rest of this writeup, I will take the point of view that all the five anomalous processes measured by
Belle at energies near the $\Upsilon(5S)$ mass~\cite{Abe:2007tk,:2008pu,Adachi:2011ji} have very little to do
with the $\Upsilon(5S)$ decays. Following~ \cite{Ali:2009es,Ali:2010pq,Ali:2009pi}, I will argue here
that the final states $\Upsilon(nS) \pi^+\pi^-$ and $h_b(mP)\pi^+\pi^-$
are the decay products of the $J^{PC}=1^{--}$ tetraquark $Y_b(10890)$, which lies in mass tantalizingly
close to the $\Upsilon(5S)$ mass. More precise experiments are needed to tell the two apart than is
the case currently.
 In the context of the $\Upsilon(nS) \pi^+\pi^-$ final states, this
was suggested in \cite{Ali:2009es,Ali:2010pq,Ali:2009pi} and the dynamical model was shown to be consistent
with the observed cross sections. Also, the measured dipion invariant mass distributions show the 
predicted scalar-and tensor-meson resonant structure. Moreover, in the tetraquark context, it is
easier to understand why the production cross sections for
 $e^+e^- \to Y_b(10890) \to \Upsilon(nS)\pi^+\pi^-$, which involves a ${}^3P \to {}^3S$ transition, and 
 for $e^+e^- \to Y_b(10890) \to h_b(mS)\pi^+\pi^-$, which involves a ${}^3P \to {}^1P$ transition,
 are comparable to each other. Detailed distributions,
including the resonant $Z_b(10610)$ and $Z_b(10650)$ effects are still being worked out
in the tetraquark picture. 
%
%
%
\section{Spectrum of bottom diquark-antidiquark states}
Much of the discussion of the tetraquark states involves the concept of diquarks (and anti-diquarks)
as effective degrees of freedom, which will be used here to calculate the mass spectra, production
and decay of the tetraquark states. In particular, four-quark configurations in the tetraquarks are
assumed not to play a dominant role. Following this,  
the mass spectrum of tetraquarks $[bq][\overline{bq^{\prime }]}$ with $q=u$, 
$d$, $s$ and $c$ can be calculated using a Hamiltonian~\cite{Drenska:2008gr}
\begin{equation}
H=2m_{\wQ}+H_{SS}^{(\wQ\wQ)}+H_{SS}^{(\wQ\bar{\wQ}\mathcal{)}}+H_{SL}+H_{LL} ,
\label{01}
\end{equation}%
where:%
\begin{eqnarray}
H_{SS}^{(\wQ\wQ)} &=&2(\mathcal{K}_{bq})_{\bar{3}}[(\mathbf{S}_{b}\cdot \mathbf{S%
}_{q})+(\mathbf{S}_{\bar{b}}\cdot \mathbf{S}_{\bar{q}})],  \nonumber \\
H_{SS}^{(\wQ\bar{\wQ}\mathcal{)}} &=&2(\mathcal{K}_{b\bar{q}})(\mathbf{S}%
_{b}\cdot \mathbf{S}_{\bar{q}}+\mathbf{S}_{\bar{b}}\cdot \mathbf{S}_{q})+2%
\mathcal{K}_{b\bar{b}}(\mathbf{S}_{b}\cdot \mathbf{S}_{\bar{b}})+2\mathcal{K}%
_{q\bar{q}}(\mathbf{S}_{q}\cdot \mathbf{S}_{\bar{q}}),  \nonumber \\
H_{SL} &=&2A_{\wQ}(\mathbf{S}_{\mathcal{Q}}\cdot \mathbf{L}+\mathbf{S}_{%
\mathcal{\bar{Q}}}\cdot \mathbf{L}),  \nonumber \\
H_{LL} &=&B_{\wQ}\frac{L_{\wQ\bar{\wQ}}(L_{\wQ\bar{\wQ}}+1)}{2}~.  \label{02}
\end{eqnarray}%
All diquarks, denoted here by $\wQ$ are assumed to be in the color triplet $(\bar{3})$, as the diquarks in the
$(6)$ representation do not show binding~\cite{Jaffe:2004ph}.
Here $m_{\wQ}$ is the constituent mass of the diquark $[bq]$, $(\mathcal{K}_{bq})_{\bar{3}%
} $ is the spin-spin interaction between the quarks inside the diquarks, $%
\mathcal{K}_{b\bar{q}}$ are the couplings ranging outside the diquark
shells, $A_{\wQ}$ is the spin-orbit coupling of diquark and $B_{\wQ}$
corresponds to the contribution of the total angular momentum of the
diquark-antidiquark system to its mass. The overall factor of $2$ is used
customarily in the literature. As the isospin-breaking effects are estimated to
be of order 5 - 8 MeV for the tetraquarks $[bq][\bar{b}\bar{q}]$~\cite{Ali:2009pi,Drenska:2008gr},
they are neglected in the mass estimates discussed below.
 
The parameters involved in the above Hamiltonian (\ref{02}) can be obtained
from the known meson and baryon masses by resorting to the constituent quark
model~\cite{De Rujula:1975ge}
\begin{equation}
H=\sum\limits_{i}m_{i}+\sum\limits_{i<j}2\mathcal{K}_{ij}(\mathbf{S}%
_{i}\cdot \mathbf{S}_{j}) , \label{Constituent-model}
\end{equation}%
where the sum runs over the hadron constituents. The coefficient $\mathcal{K}%
_{ij}$ depends on the flavour of the constituents $i$, $j$ and on the particular
colour state of the pair. The constituent quark masses and the couplings $%
\mathcal{K}_{ij}$\ for the colour singlet and anti-triplet states are given in
~\cite{Ali:2009pi}. To calculate the spin-spin interaction of the $\wQ\bar{\wQ}$ states
explicitly, one uses the non-relativistic notation~\cite{Maiani:2004vq}
$\left\vert S_{\wQ}, S_{\bar{\wQ}};~J\right\rangle $,
where $S_{\wQ}$ and $S_{\bar{\wQ}}$ are the spin of diquark and antidiquark, respectively,
and $J$ is the total angular momentum. These states are then defined in terms of the
direct product of the $2\times 2$ matrices in spinor space, $\Gamma ^{\alpha }$, which 
can be written in terms of the Pauli matrices as: 
\begin{equation}
\Gamma ^{0}=\frac{\sigma _{2}}{\sqrt{2}};~\Gamma ^{i}=\frac{1}{\sqrt{2}}%
\sigma _{2}\sigma _{i}~  ,\label{05}
\end{equation}%
which then lead to the definition such as 
$\left\vert 0_{\wQ},0_{\bar{\wQ}};~0_{J}\right\rangle =\frac{1}{2}\left( \sigma
_{2}\right) \otimes \left( \sigma _{2}\right)$. Others can be seen in~\cite{Ali:2009pi}.

 The next step is the
diagonalization of the Hamiltonian (\ref{01}) using the basis of states with
definite diquark and antidiquark spin and total angular momentum.,
There are two different possibilities~\cite{Maiani:2004vq}: 
Lowest lying $[bq][\bar{b}\bar{q}]$ states $\left( L_{\wQ\bar{\wQ}}=0\right) $ and
higher mass $[bq][\bar{b}\bar{q}]$ states $\left( L_{\wQ\bar{\wQ}}=1\right) $.
The $[bq][\bar{b}\bar{q}]$ states $\left( L_{\wQ\bar{\wQ}}=0\right)$
 can be classified in terms of the six possible states involving the 
{\it good} (spin-0) and {\it bad} (spin-1)
diquarks (here, $P$ is the parity and $C$ the charge conjugation)

\textbf{i. Two states with }$J^{PC}=0^{++}$\textbf{:}%
\begin{eqnarray}
\left\vert 0^{++}\right\rangle &=&\left\vert 0_{\wQ},0_{\bar{\wQ}%
};~0_{J}\right\rangle ;  \nonumber \\
\left\vert 0^{++\prime }\right\rangle &=&\left\vert 1_{\wQ},1_{\bar{\wQ}%
};~0_{J}\right\rangle .  \label{06}
\end{eqnarray}

\textbf{ii. Three states with }$J=1$\textbf{:}%
\begin{eqnarray}
\left\vert 1^{++}\right\rangle &=&\frac{1}{\sqrt{2}}\left( \left\vert
0_{\wQ},1_{\bar{\wQ}};~1_{J}\right\rangle +\left\vert 1_{\wQ},0_{\bar{\wQ}%
};~1_{J}\right\rangle \right) ;  \nonumber \\
\left\vert 1^{+-}\right\rangle &=&\frac{1}{\sqrt{2}}\left( \left\vert
0_{\wQ},1_{\bar{\wQ}};~1_{J}\right\rangle -\left\vert 1_{\wQ},0_{\bar{\wQ}%
};~1_{J}\right\rangle \right) ;  \nonumber \\
\left\vert 1^{+-\prime }\right\rangle &=&\left\vert 1_{\wQ},1_{\bar{\wQ}%
};~1_{J}\right\rangle .  \label{07}
\end{eqnarray}%
All these states have positive parity as both the {\it good} and {\it bad} diquarks
have positive parity and $L_{\wQ\bar{\wQ}}=0$. The difference is in the charge
conjugation quantum number, the state $\left\vert 1^{++}\right\rangle $ is even under
charge conjugation, whereas $\left\vert 1^{+-}\right\rangle $ and $%
\left\vert 1^{+-\prime }\right\rangle $ are odd.

\textbf{iii. One state with }$J^{PC}=2^{++}$\textbf{:}%
\begin{equation}
\left\vert 2^{++}\right\rangle =\left\vert 1_{\wQ},1_{\bar{\wQ}%
};~2_{J}\right\rangle .  \label{08}
\end{equation}

Keeping in view that for $L_{\wQ\bar{\wQ}}=0$ there is no spin-orbit and purely
orbital term, the Hamiltonian (\ref{01}) takes the form%
\begin{eqnarray}
H &=&2m_{\wQ}+2(\mathcal{K}_{bq})_{\bar{3}}[(\mathbf{S}_{b}\cdot \mathbf{S}%
_{q})+(\mathbf{S}_{\bar{b}}\cdot \mathbf{S}_{\bar{q}})]+2\mathcal{K}_{q\bar{q%
}}(\mathbf{S}_{q}\cdot \mathbf{S}_{\bar{q}})  \nonumber \\
&&+2(\mathcal{K}_{b\bar{q}})(\mathbf{S}_{b}\cdot \mathbf{S}_{\bar{q}}+%
\mathbf{S}_{\bar{b}}\cdot \mathbf{S}_{q})+2\mathcal{K}_{b\bar{b}}(\mathbf{S}%
_{b}\cdot \mathbf{S}_{\bar{b}}).  \label{Haml-zero}
\end{eqnarray}%
The diagonalisation of the Hamiltonian (\ref{Haml-zero}) with the states defined
above gives the eigenvalues which are needed to estimate the masses of these
states. For the $1^{++}$ and $2^{++}$ states
the Hamiltonian is diagonal with the eigenvalues \cite{Maiani:2004vq}%
\begin{eqnarray}
M\left( 1^{++}\right)  &=&2m_{\wQ}-(\mathcal{K}_{bq})_{\bar{3}}+\frac{1}{2}%
\mathcal{K}_{q\bar{q}}-\mathcal{K}_{b\bar{q}}+\frac{1}{2}\mathcal{K}_{b\bar{b%
}},  \label{09} \\
M\left( 2^{++}\right)  &=&2m_{[bq]}+(\mathcal{K}_{bq})_{\bar{3}}+\frac{1}{2}%
\mathcal{K}_{q\bar{q}}+\mathcal{K}_{b\bar{q}}+\frac{1}{2}\mathcal{K}_{b\bar{b%
}}.  \label{10}
\end{eqnarray}%
   Mass of the constituent diquark can be estimated in one of two ways:
 We take the Belle data~\cite{Abe:2007tk} as input and
identify the $Y_b(10890)$ with the lightest of the $1^{--}$ states, $Y_{[bq]}$, yielding a
diquark mass  $m_{[bq]}=5.251 \;\tn{GeV}$. This procedure is analogous to what was done
in~\cite{Maiani:2004vq}, in which the mass of the diquark $[cq]$ was fixed by using the mass of
$X(3872)$ as input, yielding $m_{[cq]}=1.933$ GeV. Instead, if we use this determination of
$m_{[cq]}$ and use the formula $m_{[bq]} =m_{[cq]}+\left( m_{b}-m_{c}\right)$, which has the virtue that the
mass difference $m_c-m_b$ is well determined, we get $m_{[bq]}=5.267\;\tn{GeV}$, yielding a difference of
$16 \;\tn{MeV}$. This can be taken as an estimate of the theoretical error on $m_{[bq]}$, which then yields
an uncertainty of about 30 MeV in the estimates of the tetraquark masses from this source alone. 
For the corresponding $0^{++}$ and $1^{+-}$ tetraquark states, there are two states each, and
hence the Hamiltonian is not diagonal. After diagonalising the $2\times 2$ matrices, the
masses of these states are obtained.

We now discuss orbital excitations with $L_{\wQ\bar{\wQ}}=1$ having both {\it good} and {\it bad}
diquarks. Concentrating on the $1^{--}$ multiplet, we recall that 
there are eight tetraquark states $[bq][\bar{b}\bar{q}]$ ($q=u, d$), and the lightest isospin doublet
is:
\begin{equation}
M_{Y_{[bq]}}^{(1)}\left( S_{\wQ}=0,~S_{\bar{\wQ}}=0,~S_{\wQ\bar{\wQ}}=0,~L_{\wQ\bar{\wQ}}=1\right) 
= m_{\left[ bq\right] }+\lambda _{1}+B_{\mathcal{\wQ}}, 
\end{equation}
and the next in mass is:
$M_{Y_{[bq]}}^{(2)}\left( S_{\wQ}=1,~S_{\bar{\wQ}}=0,~S_{\wQ\bar{\wQ}}=1,~L_{\wQ\bar{\wQ}}=1\right) 
=2m_{\left[ bq\right] }+\Delta +\lambda _{2}-2A_{\wQ}+B_{\wQ}$, and so on. 
Values of $\lambda_i(i=1,2,3) $, $A_{\wQ}$ and $B_{\wQ}$ are estimated in~\cite{Ali:2009pi}.
We identify the state $Y_b(10890)$ with  $M_{Y_{[bq]}}^{(1)}$ (in fact there are two of them, which
differ in mass from each other by about 5 - 8 MeV, including isospin-breaking). This does not fix the 
quantity $\Delta$, which  is the mass difference
of the {\it good} and the {\it bad} diquarks, i.e.
$\Delta =m_{\wQ}\left( S_{\wQ}=1\right) -m_{\wQ}\left( S_{\wQ}=0\right)$. 
 Following Jaffe and Wilczek
\cite{Jaffe:2004ph}, the value of $\Delta $ for diquark $[bq]$ is estimated as
$\Delta =202$ MeV for $q=u$, $d$, $s$ and $c$ quarks. This is another source of potential
uncertainty in estimating the tetraquark masses.
 The mass spectrum for the tetraquark states $[bq][\bar{b} \bar{q}]$
for $q=u,d$ with $J^{PC}=0^{++}, 1^{++}, 1^{+-}, 1^{--}$ and $2^{++}$
states is plotted in Fig.~\ref{fig:udspectrum-tet} 
in the isospin-symmetry limit. It is difficult to quote a theoretical error on the
masses shown, with $\pm 50$ MeV presumably a good guess.
 Other estimates of the tetraquark mass spectra in the charm and
bottom quark sectors can be seen in ~\cite{Drenska:2010kg,Ebert:2008se,Wang:2009kw}.

\subsection{Estimates of the charged $J^P=1^+$ tetraquark states}
In the tetraquark picture, one also anticipates a large number of charged states whose mass spectrum
can be calculated in an analogous fashion as for their neutral counterparts just discussed.
 We would like to propose that the two charged $J^P=1^+$ states $Z_b(10610)$ and $Z_b(10650)$
observed recently by the Belle Collaboration~\cite{Collaboration:2011gj}, and interpreted by them
as the charged bottomonium states produced in the process $\Upsilon(5S) \to  Z^\pm_b(10610) + \pi^\mp$
and $\Upsilon(5S) \to  Z^\pm_b(10650) + \pi^\mp$,
are indeed charged tetraquark states  with the quark content $Z_b^+=[b u][\bar{b}\bar{d}]$ for the positively
charged state (its charge conjugate being $Z_b^-=[\bar{b}\bar{u}][bd]$). For the present discussion, they
are produced in the decays of the $J^{PC}=1^{--}$ tetraquark $Y_b(10890)$. According to this interpretation,
the decay chains involve $Y_b(10890) \to ((Z^\pm_b(10610),Z^\pm_b(10650) +\pi^\mp \to
\Upsilon(ns) \pi^+\pi^-$. A detailed dynamical model is under development with the aim of understanding the
decay distributions in the kinematic variables available in these decays. 

\begin{figure*}[!t]
\begin{center}
\includegraphics[width=0.6\textwidth,height=0.5\textwidth]{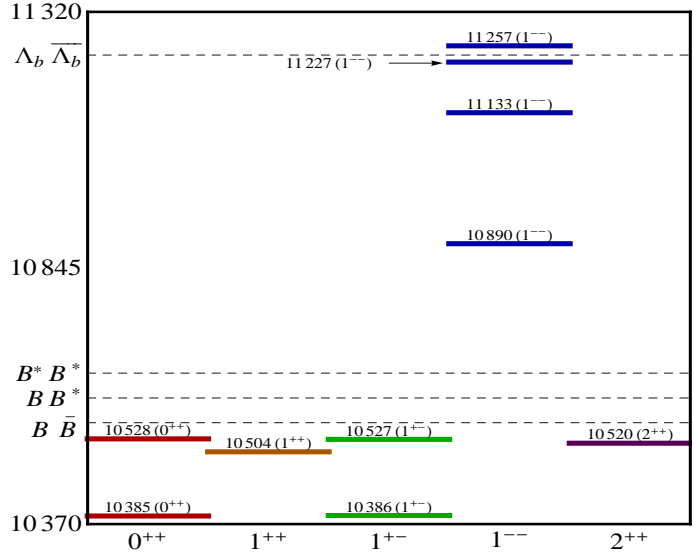}\hspace{2mm}
\vspace{-3mm}
\caption{Tetraquark mass spectrum with the valence quark content $[bq][\bar{b}\bar{q}]$ with $q=u,d$,
assuming isospin symmetry. The value 10890 is an input for the lowest $J^{PC}=1^{--}$ tetraquark state $Y_{[bq]}$. All masses are given in MeV.
   (From~\cite{Ali:2009pi}.)
\label{fig:udspectrum-tet}
}
\vspace{-3mm}
\end{center}
\end{figure*}

We have estimated the masses of the isospin partners of $Z_b(10610)$ and $Z_b(10650)$,
the two neutral $J=1$ tetraquark states, denoted as $|1^{+-}\rangle$ and $|1^{+-\prime}\rangle$.
The $2\times 2$ non-diagonal mass matrix for the neutral
 $J^{PC}=1^{+-}$ states was, however, calculated numerically for $\Delta=0$.  
If we ignore the isospin-breaking effects in the tetraquark masses, which are small,
then the charged counterparts have the masses
$M[Z_b(10610)]= 10.386$ GeV and $M[Z_b(10650)]= 10.527$ GeV, given in Fig.~\ref{fig:udspectrum-tet}. 
As $Z_b(10610)$ involves one {\it good} and one {\it bad} diquark and $Z_b(10650)$ involves two
{\it bad} diquarks, including the $\Delta$-dependent term, the non-diagonal $2\times 2$ mass matrix
gets modified to the following form
\begin{eqnarray}
 M(1^{+-}) 
&=& 2m_{\wQ} +\frac{3}{2} \Delta - \frac{\kappa_{q\bar q}+ \kappa_{b\bar b} }{2} +
				\left(\begin{array}{cc}
                                           -\frac{ \Delta }{2}-(\kappa_{bq})_{\bar 3}+ \kappa_{b\bar q}  & \kappa_{q\bar q}- \kappa_{b\bar b} \\
                                           \kappa_{q\bar q}-\kappa_{b\bar b} & \frac{ \Delta }{2} +(\kappa_{bq})_{\bar 3}- \kappa_{b\bar q}  
                                          \end{array}\right)~. 
\end{eqnarray}
The two eigenvalues can be written as $E=\pm \sqrt {x^2+y^2}$, with
 $x= \frac{ \Delta }{2} +(\kappa_{bq})_{\bar 3}- \kappa_{b\bar q}  $ and
 $y= \kappa_{q\bar q}-\kappa_{b\bar b}$, yielding
\begin{eqnarray}
 M[Z_b(10650)]=2m_{\wQ} +\frac{3}{2} \Delta - \frac{\kappa_{q\bar q}+ \kappa_{b\bar b} }{2}+
\sqrt {(\frac{ \Delta }{2} +(\kappa_{bq})_{\bar 3}- \kappa_{b\bar q} )^2+(\kappa_{q\bar q}-\kappa_{b\bar b})^2}~,\\
M[Z_b(10610)]= 2m_{\wQ} +\frac{3}{2} \Delta - \frac{\kappa_{q\bar q}+ \kappa_{b\bar b} }{2}-\sqrt {(\frac{ \Delta }{2} +(\kappa_{bq})_{\bar 3}- \kappa_{b\bar q} )^2+(\kappa_{q\bar q}-\kappa_{b\bar b})^2}~.
\end{eqnarray}
Using the  default values of the parameters~\cite{Ali:2009pi} 
\begin{eqnarray}
 m_{\wQ}= 5.251~{\rm GeV},\; (\kappa_{q\bar q})_0= 318~{\rm MeV},\;\; (\kappa_{b\bar b})_0=36 ~{\rm MeV},\;\;
 (\kappa_{b\bar q})_0=23 ~{\rm MeV},\;\; (\kappa_{b q})_3=6 ~{\rm MeV}\nonumber\\
\end{eqnarray}
\vspace*{-3mm}
we have now the following predictions for the two charged tetraquark masses
\begin{equation}
  M[Z_b(10610)]=10.637~{\rm GeV};\;\; \;M[Z_b(10650)]= 10.884~{\rm GeV}, \;\;\; {\rm with }  \;\;\Delta= 202~{\rm MeV}~.
\end{equation}
These estimates are to be compared with the masses of the $J^P=1^+$ states $Z_b(10610)$ and $Z_b(10650)$
reported by the Belle Collaboration~\cite{Collaboration:2011gj}
$M[Z_b(10610)]=(10608 \pm 2.0)$ MeV and $M[Z_b(10650)]=(10653.2 \pm 1.5)$ MeV.
They are in the right ball-park, but miss the measurements  by approximately 30 MeV and 230 MeV, respectively.
More importantly, the mass difference between the two states has been
measured precisely~\cite{Collaboration:2011gj} $M[Z_b(10650)] -  M[Z_b(10610)] \simeq 45$ MeV.
The expression for this mass difference using the Hamiltonian~(\ref{02}) is:
\begin{equation}
M[Z_b(10650)] -  M[Z_b(10610)]=2\sqrt {(\frac{ \Delta }{2} +
(\kappa_{bq})_{\bar 3}- \kappa_{b\bar q} )^2+(\kappa_{q\bar q}-\kappa_{b\bar b})^2}~.
\label{eq:massdiff}
\end{equation}
The smallest value for the mass difference (140 MeV) is obtained for $\Delta=0$, which goes up to 
  247 MeV for $\Delta= 202 {\rm MeV}$. Both are larger than the measurements. Thus, the
Belle data suggests that the Hamiltonian used here has to be augmented with an additional contribution.
As the masses of the observed states
$Z_b(10610)$ and $Z_b(10650)$ are rather close to the thresholds $M(B) + M(B^*)$ and $2 M(B^*)$,
respectively, this suggests that the threshold effects may impact on the masses and
mass differences presented here. 

\section{Tetraquark-based analysis of the processes $e^+e^- \to \Upsilon(1S) (\pi^+\pi^-,K^+K^-,\eta\pi^0)$}
\begin{figure*}[!t]
\includegraphics[width=0.3\textwidth]{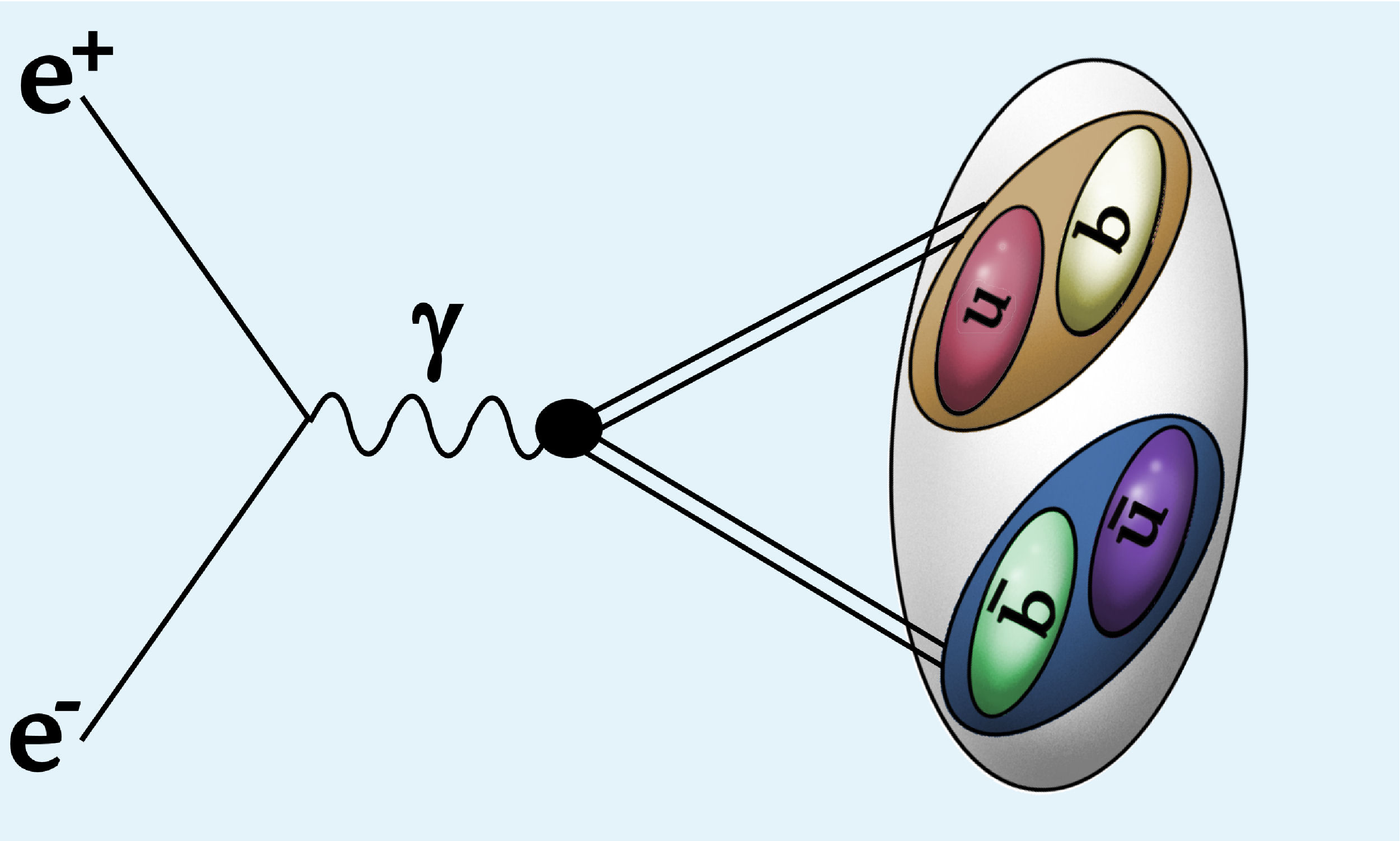}\hspace{2mm}
\includegraphics[width=0.3\textwidth]{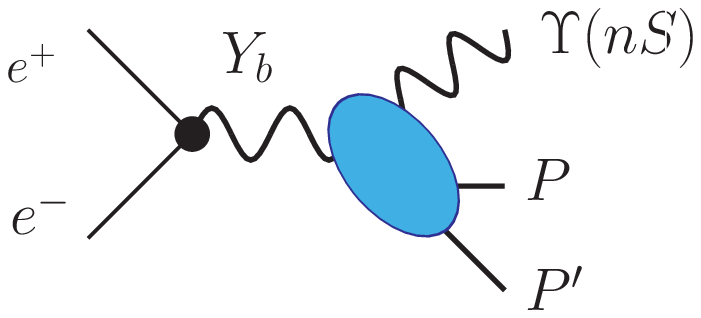}\hspace{2mm}
\includegraphics[width=0.3\textwidth]{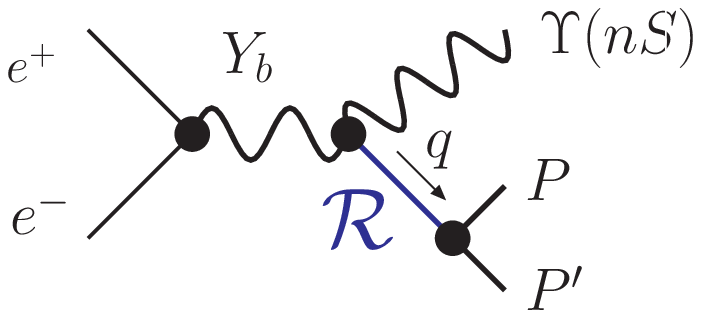}
\vspace{-3mm}
\caption{Left frame: Van Royen-Weiskopf Diagram for the production of a $J^{PC}=1^{--}$ tetraquark $Y_b$
with the quark content $[bu][\bar{b}\bar{u}]$ in the process $e^+e^- \to \gamma^* \to Y_b$.
Middle frame: Continuum contribution in the process $e^+ e^- \to Y_b \to \Upsilon(nS) P P^\prime$.
Right frame: Resonance contribution in the process $e^+ e^- \to Y_b \to \Upsilon(nS) P P^\prime$.
(Figures based on~\cite{Ali:2010pq}.)
\label{fig:tetraquark-dynamics}
}
\vspace{-3mm}
\end{figure*}

The cross sections and final state distributions for the processes
 $e^+e^- \to \Upsilon(1S) (\pi^+\pi^-,K^+K^-,\eta\pi^0)$ near the $\Upsilon(5S)$ have been presented
in the tetraquark picture in~\cite{Ali:2010pq} improving the results on the process
 $e^+e^- \to \Upsilon(1S) \pi^+\pi$ published earlier~\cite{Ali:2009es}. The distributions for the
process $e^+e^- \to \Upsilon(2S) \pi^+\pi$ calculated in~\cite{Ali:2009es} had a computational error,
which has been corrected in the meanwhile (see the Erratum in ~\cite{Ali:2009es}). These analyses are
briefly reviewed in this section.
Concentrating on the processes $e^+e^- \to \Upsilon(1S) (\pi^+\pi^-,K^+K^-,\eta\pi^0)$,
 there are essentially three important parts of the amplitude to be calculated consisting of the following:

(i) Production mechanism of the $J^{PC}=1^{--}$ vector tetraquarks in $e^+e^-$ annihilation. To that end,
we derive the equivalent of the Van-Royen-Weiskopf formula for the leptonic decay widths of
the tetraquark states $Y_{[bu]}$ and $Y_{[bd]}$ made up of a diquark and antidiquark, based on the
 diagram shown in 
Fig.~\ref{fig:tetraquark-dynamics} (left-hand frame).
\begin{equation}
\Gamma( Y_{[bu/bd]} \to e^+e^-)
=
\frac{24\alpha^2|Q_{[bu/bd]}|^2}{m_{Y_b}^4}\,
\kappa^2\left|R^{(1)}_{11}(0)\right|^2~.
\label{eq:VRW-P}
\end{equation}
Here, $Q_{[bu]}=1/3$ and $Q_{[bd]}=-2/3$ are the electric charges of the constituent diquarks of
 the $Y_{[bu]}$ and $Y_{[bd]}$,
 $\alpha$ is the fine-structure constant, the parameter $\kappa$
takes into account differing sizes of the tetraquarks compared to the
standard bottomonia, with $\kappa < 1$ anticipated,
and $| R^{(1)}_{11}(0)|^2=2.067$ GeV$^5$~\cite{Eichten:1995ch}
is the square of the derivative of the radial wave function for
$\chi_b(1P)$ taken at the origin. Hence, the leptonic widths of the
tetraquark states are estimated as  
\begin{equation}
\Gamma( Y_{[bd]} \to e^+e^-) 
= 4\,\Gamma( Y_{[bu]} \to e^+e^-) 
\approx 83\, \kappa^2\ {\rm eV}\,,
\label{eq:Ybtoee_value}
\end{equation}
which are substantially smaller than the leptonic width of the
$\Upsilon(5S)$~\cite{Amsler:2008zzb}. This is the reason why the states
$ Y_{[bd]}$ and $Y_{[bu]}$ are not easily discernible in the $R_b$-scan.
Between the two, $ Y_{[bd]}$ production dominates and should be searched for in
dedicated experiments. However, as
the decays $\Upsilon(5S) \to \Upsilon(nS)\pi^+\pi^-$ are Zweig-suppressed
in the conventional Quarkonia descriptions, and hence have small branching ratios, the signal-to-background
is much better for the discovery of the $Y_b(10890)$ 
 in the  states $\Upsilon(nS)\pi^+\pi^-$. These, in fact, are the discovery channels of
the $Y_b(10890)$~\cite{:2008pu}.

(ii) The decay amplitudes for $Y_b(10890) \to \Upsilon(1S)(\pi^+\pi^-, K^+K^-, \eta \pi^0)$ have 
non-resonant (continuum) contributions, as depicted in Fig.~\ref{fig:tetraquark-dynamics} (middle frame).
They are parametrised in terms of two {\it a priori} unknown constants $A$ and
$B$~, following~\cite{Brown:1975dz}:
\begin{eqnarray}
{\cal M}_{0}^{1C}
&=&
\frac{2A}{f_Pf_{P'}}(k_1\cdot k_2)
+ \frac{B}{f_Pf_{P'}}
\frac{3(q^0)^2k_1^0k_2^0-|\mathbf{q}|^2|\mathbf{k}|^2}{3s}
\,,\nonumber\\
{\cal M}_{0}^{2C}
&=&
- \frac{B}{f_Pf_{P'}}
\frac{|\mathbf{q}|^2|\mathbf{k}|^2}{s}
\,,
\label{eq:amp_decay-cont}
\end{eqnarray}
where the subscript $0$ denotes the $I=0$ part of the amplitudes, the superscripts 1C and 2C
correspond to the $S$- and $D$-wave continuum contributions, respectively,  
$f_{P^{(\prime)}}$ is the decay constant of $P^{(\prime)}$, and
$|\mathbf{q}|$, $k_1^0$ and $k_2^0$ are the magnitude of the three
momentum of $Y_b$ and the energies of $P$ and $P'$ in the $PP'$ rest
frame, respectively. Using SU(3) symmetry results in the relations involving the
various $I=0$ and $I=1$ amplitudes: 
${\cal M}_{0}^{1C,2C}(\ukk) = (\sqrt{3}/2)\, {\cal M}_{0}^{1C,2C}(\upp)$, 
${\cal M}_{1}^{1C,2C}(\ukk) = {\cal M}_{0}^{1C,2C}(\ukk)$
and ${\cal M}_{1}^{1C,2C}(\uep) = \sqrt{2}\, {\cal M}_{1}^{1C,2C}(\ukk)$.
We note that, in general, there is a third constant  also present in 
the non-resonant amplitudes, characterising the term depending on the
polarisation of the $Y_b$. However, being suppressed by $1/m_b$, this
is ignored.

(iii) The resonant contributions, shown in the right-hand frame of Fig.~\ref{fig:tetraquark-dynamics},
are expressed by the Breit-Wigner formula:
\begin{eqnarray}
\mathcal{M}^{R}_I
&=&
\frac{g_{R PP'}\, g_{\yb^I \Upsilon(1S) R}\, g_{\ee Y_b^0}}
{M_{PP'}^2 -m_{R}^2+ i\, m_{R}  \Gamma_{R}}\, e^{i\varphi_{R}},
\label{eq:amp_decay}
\end{eqnarray}
where $I=0$ for $R=\sigma$, $f_0$ and $f_2$, and $I=1$ for $R=a_0^0$.
The couplings for the scalar resonances $S$ are defined through the
Lagrangian  
${\cal L} = g_{SPP'} (\partial_\mu P) (\partial^\mu P')\,S
+ g_{Y_b \Upsilon(1S) S}\,Y_{b\mu} \Upsilon^\mu S$,
while those for the $f_2$ are defined via 
${\cal L} = 2\hspace{0.3mm} g_{f_2PP'}
(\partial_\mu P)(\partial_\nu P') f_2^{\mu\nu}
+ g_{Y_b \Upsilon(1S) f_2}Y_{b\mu} \Upsilon_\nu f_2^{\mu\nu}$.
The couplings $g_{R PP'}$ and $g_{\yb^I \Upsilon(1S) R}$ have mass
dimensions $-1$ and $1$, respectively. 
For the $\sigma$, $f_0$ and $a_0^0$, we adopt the Flatt\'e
model~\cite{Flatte:1976xu}  and the details can be seen in~\cite{Ali:2010pq}.

With this input, a simultaneous fit
to the binned $\upp$ data for the $M_{\pp}$ and $\cos\theta$
distributions measured by Belle at $\sqrt{s}=10.87$
GeV~\cite{Abe:2007tk} were undertaken. Normalizing the distributions by the measured cross section: 
$d\widetilde{\sigma}_{\pp}/dM_{\pi\pi}$ and 
$d\widetilde{\sigma}_{\pp}/d\cos\theta$, where
$\widetilde{\sigma}_{\pp}\equiv\sigma_{\uos\pp}/\sigma_{\uos\pp}^{\rm {Belle}}$
with $\sigma_{\uos \pp}^{\rm {Belle}}=1.61 \pm 0.16$ pb~\cite{Abe:2007tk}, 
the results are shown in Fig.~\ref{fig:spectra} (histograms) and provide
a good description of both the dipion mass spectrum and the angular distribution. 

\begin{figure*}[!t]
\includegraphics[width=0.3\textwidth]{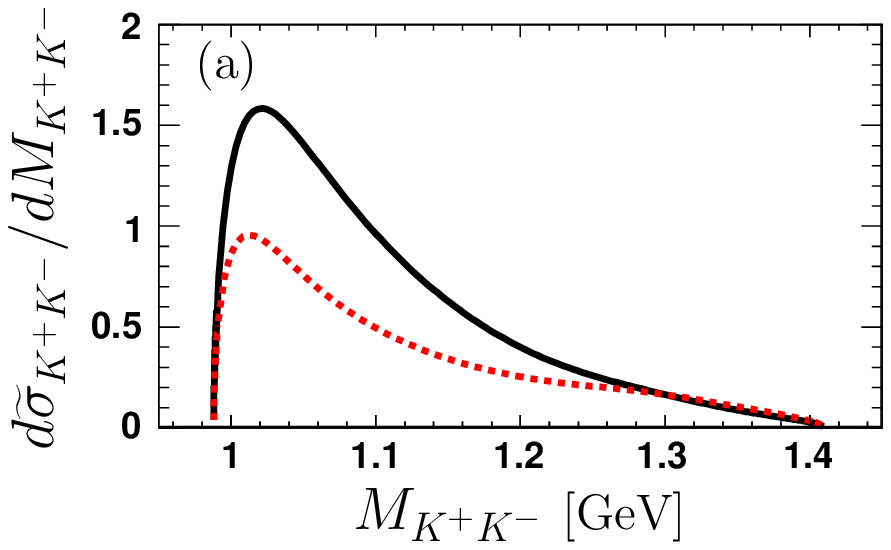}\hspace{2mm}
\includegraphics[width=0.3\textwidth]{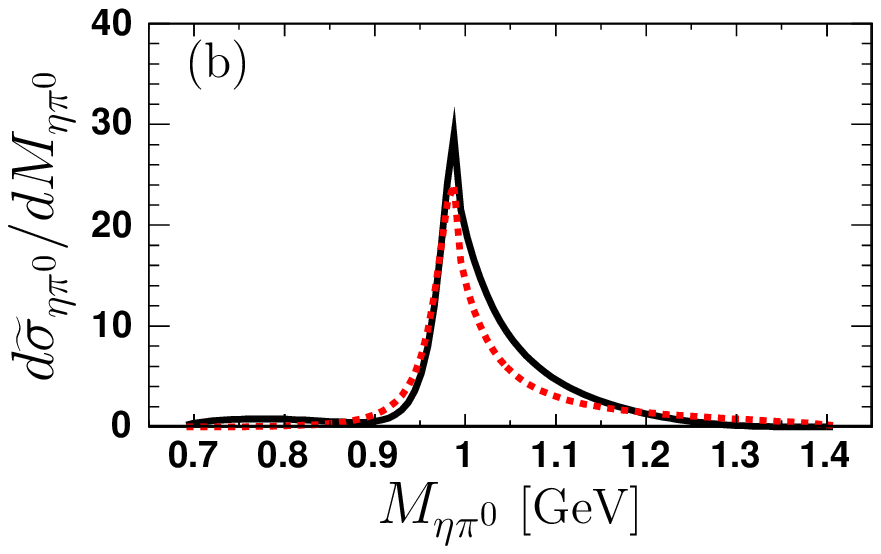}\hspace{2mm}
\includegraphics[width=0.3\textwidth]{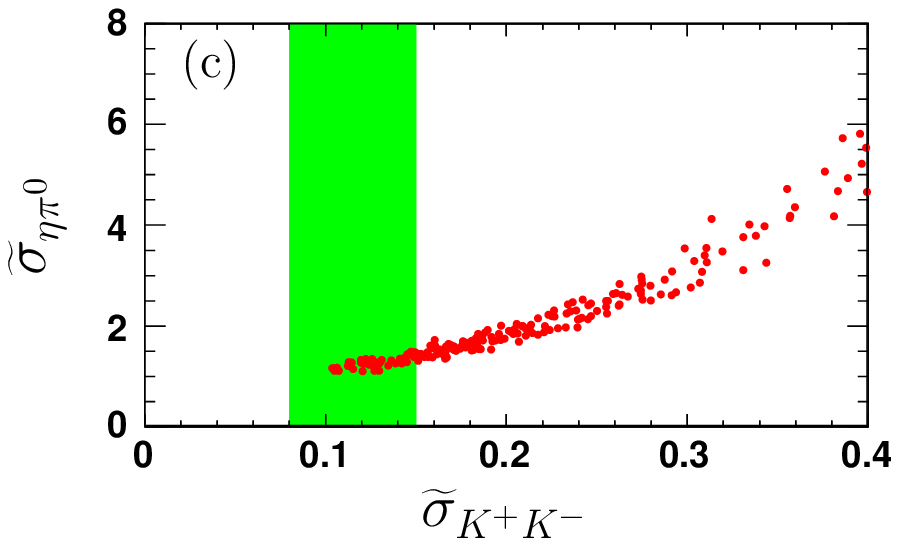}
\vspace{-3mm}
\caption{Predictions (a) of the $M_{\kk}$ distribution for
  $\ee\to Y_b \to \ukk$, (b) of the $M_{\ep}$ distribution for
  $\ee\to Y_b \to \uep$ and (c) of the correlation between the cross
  sections of $\ukk$ and $\uep$, normalized by the measured cross
  section for the $\upp$ mode. In (a) and (b), the dotted (solid)
  curves show the dimeson invariant mass spectra from the resonant
  (total) contribution. In (c), the red dots represent predictions
  from the fit solutions satisfying $\chi^2/{\rm{d.o.f.}}< 1.6$. 
  The shaded (green) band shows the current Belle measurement 
  $\widetilde{\sigma}_{K^+K^-}= 0.11^{+0.04}_{-0.03}$~\cite{Abe:2007tk}.  
(From ~\cite{Ali:2010pq}.)
\label{fig:predictions}
}
\vspace{-3mm}
\end{figure*}

\begin{figure}[t]
\centering
\resizebox{0.95\textwidth}{0.25\textwidth}{
\includegraphics[width=0.8\textwidth,height=10cm]{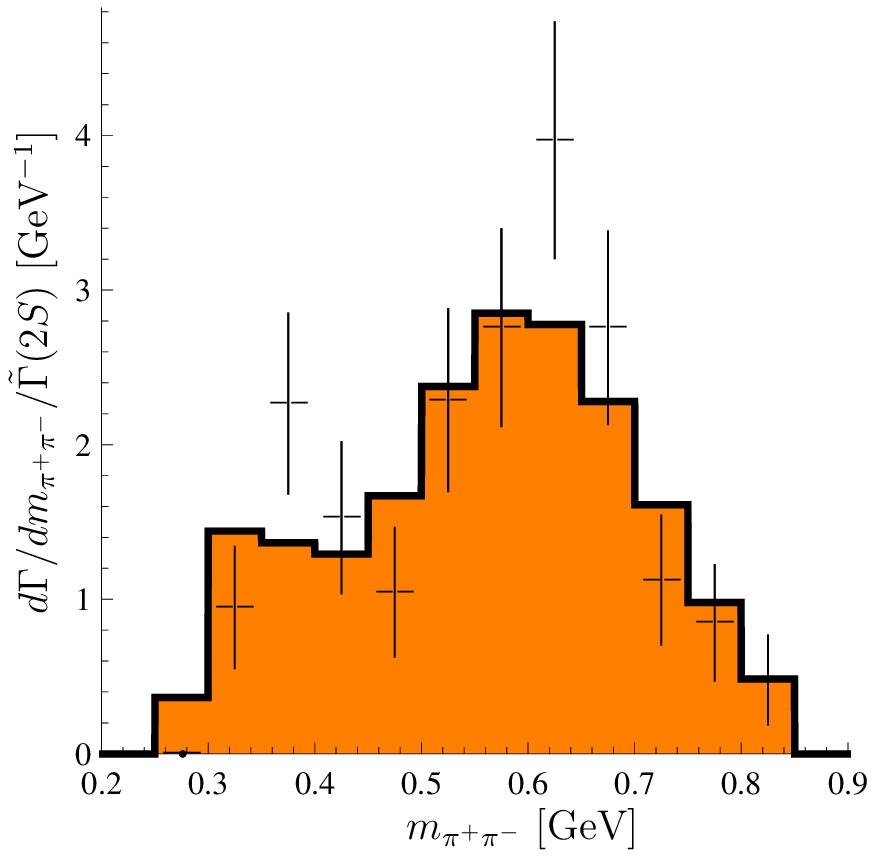}
\includegraphics[width=0.8\textwidth,height=10cm]{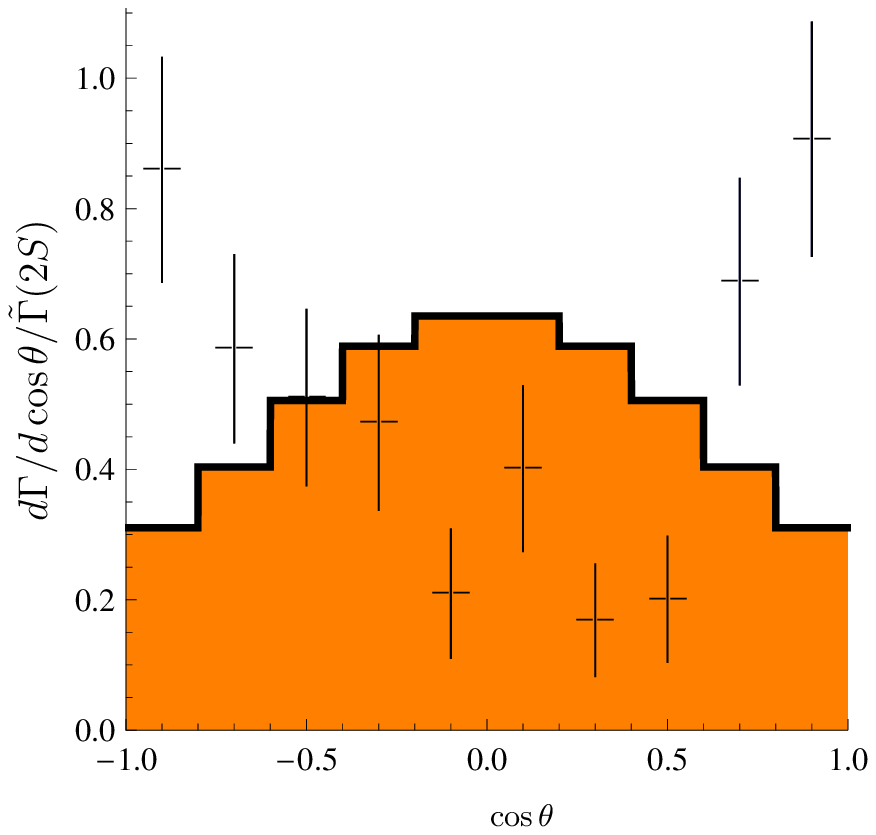}
}
\vspace*{-3mm}
\caption{Dipion invariant mass $(m_{\pi\pi})$ distribution (left-handed frame) and the $\cos \theta$
distribution (right-handed frame) measured by the Belle collaboration for the final state
 $\Upsilon(2S) \pi^+\pi^-$ ~\cite{Abe:2007tk} and the corresponding theoretical distributions
(histograms) based on the tetraquark interpretation of the $Y_B(10890)$. 
 (From ~\cite{Ali:2009es}.) 
\label{fig:spectra-2}
}
\end{figure}

The normalized $M_{K^+K^-}$ and $M_{\eta \pi^0}$ distributions
 are shown in Fig.~\ref{fig:predictions}
(a) and Fig.~\ref{fig:predictions} (b), respectively. In these
figures, the dotted (solid) curves show the dimeson invariant mass
spectra from the resonant (total) contribution. Since these spectra
are dominated by the scalars $f_0+a_0^0$ and $a_0^0$, respectively,
there is a strong correlation between the two cross sections. This is
shown in Fig.~\ref{fig:predictions} (c), where  the
normalized cross sections $\widetilde{\sigma}_{K^+K^-}$ and
$\widetilde{\sigma}_{\eta \pi^0}$ are plotted resulting from the fits (dotted 
points) which all satisfy $\chi^2/{\rm d.o.f.} < 1.6$~\cite{Ali:2010pq}.
The current Belle measurement
$\widetilde{\sigma}_{K^+K^-}=0.11^{+0.04}_{-0.03}$~\cite{Abe:2007tk}
is shown as a shaded (green) band on this figure. The tetraquark model~\cite{Ali:2010pq} is in
agreement with the Belle measurement, and 
prediction $1.0 \lesssim \widetilde{\sigma}_{\eta \pi^0} \lesssim 2.0$. 
 will be further tested as and when the cross section
$\widetilde{\sigma}_{\eta \pi^0}$ is measured.  Another important test of the
tetraquark model is~\cite{Abe:2007tk}
\begin{equation}
\frac{\sigma_{\Upsilon(1S)\kk}}
{\sigma_{\Upsilon(1S) K^0 \bar{K}^0}}
=
\frac{Q_{[bu]}^2}{Q_{[bd]}^2}
=
\frac{1}{4}\,. 
\end{equation}
This remains to be tested.
Finally, the corrected analysis~\cite{Ali:2009es} of the dipion invariant mass spectrum and the
helicity angle distribution (in $\cos \theta$) for the process $Y_b(10890) \to \Upsilon(2S)
\pi^+\pi^-$ are shown in Fig.~\ref{fig:spectra-2}, in which the normalization is given by 
the measured partial decay width $\Gamma[Y_b(10890) \to \Upsilon(2S) \pi^+\pi^-]=0.85 \pm 0.7 \pm 0.16$
MeV~\cite{:2008pu}. The dipion invariant mass spectrum is well accounted for also in this
process ($\chi^2/{\rm d.o.f.}=12.6/7  $), but not the the angular distribution $d\Gamma/d\cos \theta$.
These distributions are being reevaluated taking into account the resonances $Z_b(10610)$ and
$Z_b(10650)$.  

As a tentative summary of the tetraquark interpretation of the Belle data 
on $e^+e^- \to (\Upsilon(nS)\pi^+\pi^-$ and $e^+ e^- \to h_b(mP) \pi^+\pi^-$
is that the existing analysis are encouraging and there exists
a {\it prima facie} case of its validity. However, the missing contributions from the charged
tetraquarks in the analysis of the $e^+e^- \to (\Upsilon(nS)\pi^+\pi^-$ data
have to be incorporated and the
fits of the $e^+ e^- \to h_b(mP) \pi^+\pi^-$ data have to be undertaken 
to get a definitive answer.

I would like to thank Robert Fleischer and the organisers of the Beauty 2011 conference for a very
exciting meeting in Amsterdam. I also thank Christian Hambrock, Satoshi Mishima and Wei Wang for their help in
preparing this talk and helpful discussions.

\end{document}